\documentstyle[preprint,aps,floats]{revtex}
\def \ds { \partial \raise.3ex \hbox {\kern -.55 em/}}
\def \ps { p \raise.3ex \hbox {\kern -.55 em/}}

\def\overlay#1#2{\setbox0=\hbox{#1}\setbox1=\hbox to \wd0{\hss
#2\hss}#1\hskip -2\wd0\copy1}
\def\lsim{\mathrel{\rlap{\lower4pt\hbox{\hskip1pt$\sim$}}
    \raise1pt\hbox{$<$}}}         
\def\gsim{\mathrel{\rlap{\lower4pt\hbox{\hskip1pt$\sim$}}
    \raise1pt\hbox{$>$}}}         
\def\beq{\begin{equation}}
\def\eeq{\end{equation}}
\def\bea{\begin{eqnarray}}
\def\eea{\end{eqnarray}}
\def\bml{\begin{mathletters}}
\def\eml{\end{mathletters}}

\newcommand{\balpha}{\mbox{\boldmath $\alpha$}}
\newcommand{\brho}{\mbox{\boldmath $\rho$}}

\newcommand{\bnabla}{\mbox{\boldmath $\nabla$}}

\newcommand{\btau}{\mbox{\boldmath$\tau$}}
\newcommand{\bgamma}{\mbox{\boldmath{$\gamma$}}}
\newcommand{\bomega}{\mbox{\boldmath{$\omega$}}}
\newcommand{\bSigma}{\mbox{\boldmath{$\Sigma$}}}
\begin{document}
\preprint{TAUP-2310-95, IFT-P.003/96, nucl-th/9601027}
\title{Relativistic deformed mean-field calculation of binding energy
differences of mirror nuclei}
\author{W. Koepf\\
School of Physics and Astronomy, Tel Aviv University,
69978 Ramat Aviv, Israel}
\author{G. Krein \\
Instituto de F\'{\i}sica Te\'orica, Universidade Estadual Paulista, 
Rua Pamplona 145\\
01405-900 S\~ao Paulo-SP, Brazil}
\author{L.A. Barreiro\\
Instituto de F\'{\i}sica, Universidade de S\~ao Paulo, Caixa Postal 66318 \\
05389-970 S\~ao Paulo-SP, Brazil}
\date{January 16, 1996}
\maketitle

\begin{abstract}
Binding energy differences of mirror nuclei for A=15, 17, 27, 29, 31, 33, 39 
and 41 are calculated in the framework of relativistic deformed mean-field 
theory.  The spatial components of the vector meson fields and the photon are 
fully taken into account in a self-consistent manner. The calculated binding 
energy differences are systematically smaller than the experimental values and 
lend support to the existency of the Okamoto--Nolen-Schiffer anomaly found 
decades ago in nonrelativistic calculations. For the majority of the nuclei 
studied, however, the results are such that the anomaly is significantly 
smaller than the one obtained within state-of-the-art nonrelativistic 
calculations. 
\end{abstract}
\newpage
\hspace*{-\parindent}{\bf 1. Introduction}.\hspace*{\parindent}
Conventional nuclear theory, based on the Schr\"odinger equation, has
difficulties in explaining the binding energy differences of mirror 
nuclei~\cite{{NS},{Okam}}. The major contribution to the energy difference
comes from the Coulomb force, 
and a variety of isospin breaking effects provide small
contributions. The difficulty in reproducing the experimental
values is known in the literature as the Okamoto-Nolen-Schiffer anomaly 
(ONSA). Several nuclear structure effects such as correlations, core 
polarization and isospin mixing have been invoked to resolve the anomaly 
without success\cite{Auerb}. 

Since Okamoto and Pask~\cite{OkamPash} and 
Negele~\cite{Neg} suggested that the discrepancy could be due to a small charge
symmetry breaking component in the nuclear force, a variety of calculations
have been performed following this suggestion, and a widespread consensus 
has emerged that the anomaly can eventually be explained by a charge 
symmetry violation in the  nucleon-nucleon interaction~\cite{GAMrev1}. 
In particular,
class III (pp-nn) and class IV (pn)~\cite{HM} charge symmetry breaking (CSB) 
forces can affect the binding energy differences of mirror 
nuclei~\cite{III-IV-BE}, with the effects of the $\rho^0-\omega$ mixing 
being responsible for the bulk of the anomaly~\cite{{BI},{jap}}. However,
although the situation regarding the resolution of the ONSA in terms of the
$\rho^0-\omega$ mixing interaction looks very satisfactory, there have been
several recent discussions in the literature~\cite{momdep} suggesting that 
the amplitude of the mixing is strongly momentum dependent such that the
resulting CSB potential is small at internucleon separations relevant for 
the anomaly. 

In view of such results, it is important to investigate the issue of the 
binding energy differences of mirror nuclei also
in the framework of a relativistic
nuclear structure model~\cite{wal,hs,SW.86}, where the cancellation of 
very strong scalar and vector potentials is responsible for the binding. 
Particularly important, in that realm, is the full inclusion of the effects of 
the polarization of the nuclear core due to an extra particle or hole
that are mediated through those strong fields.
In the language of relativistic mean-field theory, this requires that the
spatial components of the vector fields and the photon are taken into
account in a self-consistent manner. The latter are usually neglected in
investigations in that realm, and they actually vanish if the system 
under consideration is invariant under the operation of time-reversal,
which is the case, for instance, for the ground-state of even-even nuclei.
However, where time-reversal invariance is violated, as, for example, in
odd mass or rotating nuclei, the polarization effects generated by the
space-like components of the vector fields turned out to be very important.
In the framework of relativistic mean-field theory,
this was observed in studies of the magnetic moments in odd mass
nuclei~\cite{hofmann} and in an investigation of the moments of inertia in 
rotating nuclei~\cite{koenig}.

It is well known that the major contribution to the binding energy 
difference of mirror nuclei arises from the Coulomb force, and it 
is therefore essential to take the effects of the electromagnetic 
interaction into account as accurately as possible. In this work, we evaluate
the Coulomb field in the nuclei under consideration self-consistently,
and we fully include the effects of the nucleon's anomalous magnetic 
moments by means of a tensor-coupling to the electromagnetic
field strength tensor. Furthermore, we include the Fock exchange 
contribution to the Coulomb energy in the Slater approximation 
\cite{slater}, whereas all other fields are treated in the Hartree
approximation only.

Also, we do not imply the constraint of spherical symmetry, which would
greatly simplify the numerical evaluation, 
but we allow the self-consistent solutions of the respective field 
equations, the Dirac equation for the nucleon spinors and inhomogeneous 
Klein-Gordon equations for the mesonic and the Coulomb fields, to be axially 
deformed. 

By calculating the ground-state binding energies of various light odd mass 
nuclei in that manner, we essentially perform a state-of-the-art evaluation 
of the Okamoto-Nolen-Schiffer anomaly in the framework of the modern
relativistic mean-field theory. Many contributions to the anomaly, that in 
other works in that realm had been accounted for only perturbatively, are 
fully included in our description. Amongst them are the core polarization,
i.e., the isospin impurity of the $N=Z$ nuclear core\cite{core}, 
dynamical effects 
of the proton-neutron mass difference \cite{dyna}, contributions of the
electromagnetic spin-orbit interaction and the Darwin term \cite{NS,auer}, 
the so-called Thomas-Ehrman shift \cite{thom} and the Coulomb exchange 
term \cite{NS,dyna}.

In detail,
in this paper we calculate the binding energies of mirror nuclei with A=15, 
17, 27, 29, 31, 33, 39 and 41 using a deformed relativistic mean-field 
model. The first published calculation of binding energy differences 
of mirror nuclei using a relativistic nuclear model is by Nedjadi and 
Rook~\cite{NedRoo}. In their calculation, however, the nuclear structure
was described in single-particle approximation in terms of a Dirac equation
with spherical scalar and vector Woods-Saxon potentials, and no
self-consistency between the potentials and the respective solutions of
the Dirac equation was enforced. In a recent publication~\cite{BGK},
self-consistency between the potentials and the Dirac sources was achieved,
but again while assuming spherical symmetry, and also the effects of the
$\rho^0-\omega$ mixing interaction on the binding energies of mirror nuclei 
in the region of A=16 and A=40 were calculated within this approach. Here,
we evaluate the binding energy  differences employing the relativistic
state-of-the-art model outlined above.
 
\vspace{0.5cm}
\hspace*{-\parindent}{\bf 2. The model}.\hspace*{\parindent}
The starting point of any investigation in the realm of relativistic 
mean-field theory is the local Lagrangian density,
\bea
{\cal L}&=& \bar \psi_N \left(
           i\rlap{/}\partial
           -g_\sigma \sigma
           -g_\omega \rlap{/}\omega
           -g_\rho \brho\llap{/}\btau
           -e_N A\llap{/}
           -\kappa_N \sigma_{\mu\nu}F^{\mu\nu}
           -M_N
           \right) \psi_N
   \nonumber\\
&&~~~+{1\over2}\partial_\mu\sigma\partial^\mu\sigma - U(\sigma)
-{1\over 4}\Omega_{\mu\nu}\Omega^{\mu\nu} 
+{1\over2}m_\omega^2\omega_\mu\omega^\mu
\nonumber\\ 
&&~~~-{1\over4}{\bf R}_{\mu\nu}{\bf R}^{\mu\nu} +
{1\over2}m_\rho^2 \brho_\mu \brho^\mu - 
{1\over 4}F_{\mu\nu}F^{\mu\nu},
\label{B1}
\eea
consisting of nucleonic, mesonic and electromagnetic fields. The Dirac spinor
nucleons ($\psi_N$) couple to the isoscalar-scalar $\sigma$-meson, the
isoscalar-vector $\omega$-meson, the isovector-vector $\rho$-meson and the 
electromagnetic field, and $g_\sigma$, $g_\omega$, $g_\rho$ and $e_N$ 
are the respective coupling constants. 
The field tensors for the vector mesons and the photon field are:
\begin{equation}
\Omega_{\mu\nu} = \partial_\mu\omega_\nu -\partial_\nu\omega_\mu,\hspace{0.3cm}
{\bf R}_{\mu\nu} = \partial_\mu\brho_\nu -\partial_\nu\brho_\mu,\hspace{0.3cm}
F_{\mu\nu} = \partial_\mu{\bf A}_\nu -\partial_\nu{\bf A}_\mu.
\end{equation}        
The $\sigma$-meson has a nonlinear
self-coupling given by the potential $U(\sigma)$
\beq 
   U(\sigma) = {1\over 2}m^2_\sigma\sigma^2 +
     {1\over3}g_2\sigma^3+{1\over4}g_3\sigma^4 ,
\label{B3}
\eeq
which was found found to be important, in particular, for an
adequate description of nuclear surface and compression properties
\cite{boguta}. Note that in addition to the standard description \cite{SW.86}, 
in Eq.~(\ref{B1}) also the tensor coupling of the nucleon's anomalous 
magnetic moments, $\kappa_p=1.793$ and $\kappa_n=-1.913$, with the 
electromagnetic field has been included.

In the mean-field approximation, the meson and photon fields are treated 
classically, and the variational principle leads to time-independent
inhomogeneous Klein-Gordon equations for the mesonic fields with source 
terms involving the various nucleonic densities and currents:
\bml                   
\bea
\left\{ -\Delta+m^2_\sigma \right\}\;
   \sigma({\bf r})~&=&-~g_\sigma  
   \left[\rho_s^p({\bf r})+\rho_s^n({\bf r})\right]
    - g_2 \sigma^2({\bf r}) - g_3 \sigma^3({\bf r}) ,
\label{B4a} \\
\left\{-\Delta+m^2_\omega\right\}
   \omega_0({\bf r})&=&  ~g_\omega
   \left[\rho_v^p({\bf r})+\rho_v^n({\bf r})\right] ,
\label{B4b} \\
\left\{ -\Delta+m^2_\omega\right\} 
   \bomega({\bf r})&=&  ~g_\omega
   \left[{\bf j}_v^p({\bf r})+{\bf j}_v^n({\bf r})\right] ,
\label{B4c} \\
\left\{-\Delta+m^2_\rho\right\}
   \rho_0({\bf r})&=&  ~g_\rho
   \left[\rho_v^p({\bf r})-\rho_v^n({\bf r}\right] ,
\label{B4d} \\
\left\{ -\Delta+m^2_\rho\right\} 
   \brho({\bf r})&=&  ~g_\rho
   \left[{\bf j}_v^p({\bf r})-{\bf j}_v^n({\bf r})\right] ,
\label{B4e} \\
-\Delta\,A_0({\bf r}) 
   &=&  ~e\,\rho_v^p({\bf r}) + 2i\,\bnabla\cdot
   \left[\kappa_p\,{\bf j}_s^p({\bf r})+
   \kappa_n\,{\,\bf j}_s^n({\bf r})\right] ,
\label{B4f} \\
-\Delta\,{\bf A}({\bf r}) 
   &=&  ~e\,{\bf j}_v^p({\bf r})~+~2\,\bnabla\times
   \left[\kappa_p\,{\bf j}_\Sigma^p({\bf r})+
   \kappa_n\,{\bf j}_\Sigma^n({\bf r})\right] .
\label{B4g}
\eea
\eml                 
The corresponding source terms are:
\bml                   
\bea
&& \hspace{1.75cm}\rho_s^N = \sum_{i=1}^{N,Z}\bar\psi_i\,\psi_i, \hspace{0.5cm}
   \rho_v^N = \sum_{i=1}^{N,Z}\psi^+_i \psi_i,\label{B5a}\\
&& {\bf j}_v^N = \sum_{i=1}^{N,Z}\bar\psi_i\,\bgamma\,\psi_i,
\hspace{0.5cm} 
{\bf j}_s^N = \sum_{i=1}^{N,Z}\psi_i^+ \bgamma\,\psi_i ,
\hspace{0.5cm} 
{\bf j}_\Sigma^N = \sum_{i=1}^{N,Z}\bar\psi_i\,\bSigma\,\psi_i ,
\label{B5b}
\eea
\eml                 
where the sums run over the valence nucleons only. As usual, we
neglect the contributions from negative energy states
({\it no-sea} approximation).

In this investigation, we limit ourselves to the Hartree approximation.
The only exception is the Coulomb interaction for which a Fock exchange
contribution is included in the Slater approximation \cite{slater}.
Then, the Dirac equation for the nucleons can be written as:
\beq
   \Bigl\{ \balpha\left(-i\nabla-{\bf V}({\bf r})\right)
   ~+~V_0({\bf r})~+~
   \beta\left[M_N+S({\bf r})\right] \Bigr\} \psi_i~=~\epsilon_i \psi_i.
\label{B6}
\eeq
It contains an attractive scalar potential,
\beq
   S({\bf r})~=~g_\sigma \sigma({\bf r})
               ~-~2\,\kappa_N\,\bSigma\cdot
               \left[\bnabla\times{\bf A}({\bf r})\right],
\label{B7}
\eeq
a repulsive vector potential,
\beq
   V_0({\bf r})~=~g_\omega \omega_0({\bf r})
            +g_\rho \tau_3 \rho_0({\bf r})
            +e_N A_0({\bf r}),
\label{B8}
\eeq
and a magnetic potential,
\beq
   {\bf V}({\bf r})~=~g_\omega \bomega({\bf r})
            +g_\rho \tau_3 \brho({\bf r})
            +e_N {\bf A}({\bf r})
            -2i\,\kappa_N\,\beta\,\bnabla A_0({\bf r}),
\label{B9}
\eeq
which lifts the degeneracy between nucleonic states related by 
time-reversal.

Since we are considering nuclei with odd particle numbers,
time-reversal invariance is broken in our calculations, and we have to 
take into account also the nuclear currents of Eq.~(\ref{B5b}). The latter 
are usually neglected in investigations in that realm, although they have 
proven to be important, for instance, for a successful description of the
magnetic moments of odd mass nuclei~\cite{hofmann} 
as well as the moments of inertia in rotating nuclei \cite{koenig} --
where time-reversal is broken by the Coriolis field -- in a description in
relativistic self-consistent cranking theory, as
developed in Ref.~\cite{koepf}.
Those currents are the sources for the space-like
components of the vector $\bomega({\bf r})$, $\brho({\bf r})$ and 
${\bf A}({\bf r})$ fields -- see, e.g., Eqs.~(\ref{B4c}), ~(\ref{B4e}) 
and ~(\ref{B4g}) -- which, in turn, give rise to polarization effects 
in the Dirac spinors through the magnetic potential ${\bf V}({\bf r})$
of Eq.~(\ref{B9}). As the latter destroys the degeneracy between nucleonic 
states related via time-reversal, for instance, for odd mass nuclei as 
studied here, the odd nucleon polarizes the even-even nuclear core.
This effect is commonly referred to as {\it nuclear magnetism\/}.

These equations are solved self-consistently following the method based on 
an expansion in terms of eigenfunctions of an axially symmetric deformed 
harmonic oscillator, as developed by the Munich group \cite{koepf,thimet}, 
and the basis is truncated such that reliable convergence is achieved.
This expansion technique introduces additional basis parameters which are 
optimally chosen so as to get fast convergence (for details see Ref.
\cite{thimet}). In detail, we fix the oscillator length, $b_0 =
\sqrt{\hbar/M_N{\omega}_0}$, corresponding to $\hbar\omega_0 =
41A^{-1/3}$ (for bosons $b_B = b_0/\sqrt{2}$), where $M_N$ is the nucleon
mass.

Pairing is not included in our investigation because there is
little known about this correlation in these light nuclei under 
consideration. Also, as we are only interested in binding 
energy differences between mirror nuclei, pairing effects cancel
out almost entirely.

We use the recently proposed non-linear Lagrangian parameter set 
NL-SH \cite{nlsh1} which has been shown to yield excellent results for
ground-state binding energies, charge and neutron radii of spherical
as well as deformed nuclei on both sides of the stability line 
\cite{nlsh2}. The exact values for the various parameters in the
Lagrangian of Eq.~(\ref{B1}) are given in Table 1.

\vspace{0.3cm}
\noindent
Table 1. Mass parameters and coupling constants for the non-linear
parameter set NL-SH.
\vspace{3.0mm}
\begin{center}
\begin{tabular}{l@{~~~~~~~~}l} 
\hline\hline
Masses & Coupling constants \\
\hline
$m_\sigma=526.059$ MeV & $g_\sigma=~10.444$ \\
                       & $g_2     = -6.910$ fm$^{-1}$ \\
                       & $g_3     =-15.834$ \\
$m_\omega=783.000$ MeV & $g_\omega=~12.945$ \\
$m_\rho  =763.000$ MeV & $g_\rho  =~ 4.383$ \\
\hline\hline
\end{tabular}
\end{center}
\vspace{3.0mm}

Note that in our numerical calculations, the total angular momentum, $j$, is
not a good quantum number -- due to the, in general, nonvanishing axial 
deformation of the nucleon source terms and mesonic fields -- but
only its projection onto the symmetry axis, $m_j$, and we are thus evaluating
the respective binding energies in the nucleus's intrinsic reference frame.  
We could improve on that approximation by projecting the corresponding 
solutions onto good angular momentum.  However, as we are only interested in 
binding energy differences, the uncertainties that are associated with the 
aforementioned approximation are expected to be small.

\vspace{0.5cm}
\hspace*{-\parindent}{\bf 3. Numerical results}.\hspace*{\parindent}
Table 2 summarizes the results for the binding energy differences for 
various nuclei between A=15 and A=41. The experimental values are presented in
column EXP. In this table we also list the nonrelativistic results obtained
by Sato~\cite{sato} in the framework of the density matrix expansion (DME) 
method and Skyrme II Hartree-Fock calculations (SkII), respectively.  
The results of the present
calculation are presented in column REL. Columns $\Delta_{\rm DME}$,
$\Delta_{\rm SkII}$, and $\Delta_{\rm REL}$ are the differences between
the experimental values and the respective results of the nonrelativistic (DME,
SkII) and relativistic (REL) theoretical calculations. 

\vspace{0.3cm}
\noindent
Table 2. Binding energy differences in keV. Columns EXP are the experimental
values, DME and SkII are the results of the nonrelativistic calculations by 
Sato~\cite{sato}, and REL refers to the present relativistic calculation. 
Columns $\Delta_{\ldots}$ are the respective differences between theory and 
experiment.
\vspace{3.0mm}
\begin{center}
\begin{tabular}{c@{~~}c@{~~}|@{~~}c@{~~}|
                @{~~}c@{~~}@{~~}c@{~~}@{~~}c@{~~}@{~~}c@{~~}|
                @{~~}c@{~~}@{~~}c@{~~}@{~~}c@{~~}@{~~}c@{~~}}
\hline\hline
A & State & EXP & DME & $\Delta_{\rm DME}$ & SkII & $\Delta_{\rm SkII}$ 
                & REL & $\Delta_{\rm REL}$ \\ 
\hline
15 & $1p_{1/2}^{-1}$& 3560 & 3180 & 380 & 3270 & 290 & 3465 &  95\\
   & $1p_{3/2}^{-1}$& 3460 & 3215 & 245 & 3270 & 190 & 3239 & 221\\
17 & $1d_{5/2}^{}$  & 3500 & 3200 & 300 & 3305 & 195 & 3421 &  79\\
27 & $1d_{5/2}^{-1}$& 5610 & 5130 & 480 & 5115 & 495 & 5059 & 551\\
29 & $2s_{1/2}$     & 5700 & 5415 & 285 & 5465 & 235 & 5666 &  34\\
31 & $2s_{1/2}^{-1}$& 6250 & 5710 & 540 & 5685 & 565 & 6211 &  39\\
33 & $1d_{3/2}$     & 6350 & 5990 & 360 & 6070 & 280 & 6137 & 213\\
39 & $1d_{3/2}^{-1}$& 7430 & 6895 & 535 & 7000 & 430 & 7026 & 404\\
41 & $1f_{7/2}$     & 7230 & 6790 & 440 & 6875 & 355 & 6826 & 404\\
\hline\hline
\end{tabular}
\end{center}
\vspace{3.0mm}

The first and most concrete conclusion, one can draw from Table 2, is that --
similarly to the DME and SkII calculations -- the relativistic results for the 
binding energy differences are systematically smaller than the experimental 
values.  Second, for the  nuclei with A = 15($1p_{1/2}^{-1}$), 17, 29, 31, 33 
and 39, the relativistic 
results are significantly closer to experiment than the nonrelativistic ones, 
and for A = 15($1p_{3/2}^{-1}$) and 41, 
the deviation of the relativistic calculation from
experiment is between the SKII and the DME results. Third,
it is interesting to observe that for A=15 the experiment-theory difference,
$\Delta_{\rm REL}$, is larger for the $1p_{3/2}^{-1}$ than for the 
$1p_{1/2}^{-1}$ state, contrary to the corresponding nonrelativistic DME and 
SkII results. Note, also, that the relativistic results go 
in the same direction as the experimental values, i.e. REL(3/2) $>$ REL(1/2) 
and EXP(3/2) $>$ EXP(1/2). This feature was already observed in the
previous relativistic calculation of Ref.~\cite{BGK}, and it hints the
superiority of the relativistic framework
in the description of the nuclear spin-orbit interaction,
as discussed, for instance, in Ref.~\cite{koepf1}.
Fourth, only for nuclei with A=27, the differences between theory and
experiment are larger in the relativistic calculation
than in the nonrelativistic ones.  The corresponding numerical calculations
converge actually very slowly, indicating that other degrees of freedom, e.g.,
triaxial deformation or multi-nucleon correlations, play an important role
for those nuclei.

\vspace{0.5cm}
\hspace*{-\parindent}{\bf 3. Conclusions and future
perspectives}.\hspace*{\parindent}
Although our calculation still leaves room for improvement, such as the 
inclusion of
exchange effects or projection onto good angular momentum, it
is, however, fair to conclude that the results support, within the context of 
a relativistic mean-field description, the existence of the ONSA. However, the
relativistic results have a tendency to be closer to experiment than the
corresponding state-of-the-art nonrelativistic ones, i.e., the respective
ONS anomaly is smaller in the relativistic description.

Nonrelativistic and relativistic calculations have shown that inclusion
of $\rho^0-\omega$ mixing in the NN interaction can resolve the ONSA in
a satisfactory way. However, in view of several recent discussions suggesting 
that the contribution of the $\rho^0-\omega$ mixing to the ONSA is strongly 
suppressed, it is extremely important to investigate alternative explanations
of the anomaly. Various current
unconventional ideas on the origin of the anomaly will be checked to high
precision in the framework presented here~\cite{future}. Amongst them are a
isovector component in the coupling of the otherwise isoscalar $\sigma$-meson,
as suggested by Saito and Thomas~\cite{saito}, i.e.  a small difference in the
couplings of the proton and the neutron to the scalar $\sigma$-field. A similar
effect, which will be incorporated into the present model, is the 
modification of the in-medium up-down quark condensates, which, in
turn, implies a change of the neutron-proton mass difference in the 
nucleus, as proposed by Henley and Krein~\cite{HK}. Another hypothesis,
which can be evaluated very precisely in the model presented here, is a 
possible isospin violation in various meson-nucleon coupling constants, 
as suggested independently by Dmitra\u sinovi\' c and Pollock~\cite{pollock} 
as well as Gardner et al.~\cite{gardner}. All these effects can, in principle,
arise through the small mass difference of the up- and down-quarks in the 
nucleon. 

\vspace{0.5cm}
\hspace*{-\parindent}{\bf Acknowledgments}. \hspace*{\parindent}
W.K. wishes to thank N.~Auerbach for many useful discussions. This work was
partially supported by CNPq and FAPESP (Brazilian agencies), and
by the MINERVA Foundation of the Federal Republic of Germany.

\end{document}